# Learning to reconstruct the bubble distribution with conductivity maps using Invertible Neural Networks and Error Diffusion

N. Kumar[1*], L. Krause[2,3], T. Wondrak[3], S. Eckert[3], K. Eckert[2,3], S. Gumhold[1]

[1]Chair of Computer Graphics and Visualization, TU Dresden, Germany

[2]Institute of Process Engineering and Environmental Technology, TU Dresden, Germany

[3]Institute of Fluid Dynamics, Helmholtz-Zentrum Dresden-Rossendorf, Dresden, Germany

[*]Email: nishant.kumar@tu-dresden.de

## ABSTRACT

*Electrolysis is crucial for eco-friendly hydrogen production, but gas bubbles generated during the process hinder reactions, reduce cell efficiency, and increase energy consumption. Additionally, these gas bubbles cause changes in the conductivity inside the cell, resulting in corresponding variations in the induced magnetic field around the cell. Therefore, measuring these gas bubble-induced magnetic field fluctuations using external magnetic sensors and solving the inverse problem of Biot-Savart's Law allows for estimating the conductivity in the cell and, thus, bubble size and location. However, determining high-resolution conductivity maps from only a few induced magnetic field measurements is an ill-posed inverse problem. To overcome this, we exploit Invertible Neural Networks (INNs) to reconstruct the conductivity field. Our qualitative results and quantitative evaluation using random error diffusion show that INN achieves far superior performance compared to Tikhonov regularization.*

## 1   INTRODUCTION

The increasing demand for clean energy has driven extensive research on electrolysis for hydrogen production, offering advantages like zero greenhouse gas emissions, energy storage capabilities, and a promising pathway towards reducing the carbon footprint (Capurso et al., 2022). However, the efficiency of electrolysis is limited by the formation of gas bubbles that impede the reaction and block electric currents, thereby decreasing the efficiency of the electrolysis cell for producing hydrogen (Angulo et al., 2020). Therefore, the detection of both bubble sizes and gas distribution, as well as the control of the bubble formation, is crucial for ensuring the safety and sustainability of hydrogen production via electrolysis.

Locating bubbles in electrolysis cells is difficult as the electrolyzer structures are typically non-transparent. However, an easy and non-invasive approach to address this problem is to use externally placed magnetic field sensors to measure bubble-induced fluctuations. However, the availability of only low-resolution magnetic field measurements outside the cell, coupled with the high-resolution current distribution inside the cell necessary to provide bubble information creates an ill-posed inverse problem for bubble detection. Additionally, bubble growth and detachment are governed by a complex interplay of various forces, such as buoyancy, hydrodynamic and electrostatic forces (Hossain et al., 2020), while measurement errors due to sensor noise add to the challenge of bubble detection.

Contactless Inductive Flow Tomography (CIFT), pioneered by (Stefani and Gerbeth, 1999), enables the reconstruction of flow fields in conducting fluids by utilizing Tikhonov regularization. The technique estimates induced electric and magnetic fields resulting from fluid motion under an applied magnetic field, with the measurements taken from magnetic sensors placed on the external walls of the fluid volume. However, in our current tomography configuration, we do not induce current through an external magnetic field, and the limited number of available sensors poses an added problem in achieving a satisfactory reconstruction of the high-dimensional current distribution.

Deep Neural Networks (DNNs) offer a data-driven approach to reconstruct the current distribution inside an electrolysis cell based on external magnetic field measurements, thereby capturing complex relationships between the two. A method called Network Tikhonov (NETT) (Li et al., 2020) combines DNNs with Tikhonov regularization, where a regularization parameter α plays a crucial role in balancing data fidelity and regularization terms. However, selecting an appropriate α can be challenging, as it impacts the quality of outcomes and often relies on heuristic assumptions (Hanke, 1996).

We applied Invertible Neural Networks (INNs) to reconstruct the current distribution in 2D from one-dimensional magnetic field measurements, aiming to capture 200 times more features in the output compared to the input space. However, the INN struggles to generalize due to limited and low-resolution magnetic field data, resulting in poor reconstruction or significant overfitting. The lack of information hampers the performance despite adding latent variables to match dimensionality. We also explored Fourier analysis solutions, as suggested by (Roth et al., 1989), but as the authors pointed out, it proved insufficient due to the high sensor distance from the current distribution and noise in sensor readings.

To address the limitation of reconstructing high-resolution current distribution with limited magnetic sensors, we explored an alternative approach based on lower-resolution binary conductivity maps. These discrete maps represent non-conducting void fractions as zeros, indicating the presence of bubbles. A cluster of zeros can indicate either the existence of large bubbles or a cluster of small bubbles, enabling us to estimate the bubble distribution and their sizes. We define the conductivity map as $x \in \mathbb{R}^N$ and the magnetic field measurements as $y \in \mathbb{R}^M$ where $N > M$ such that the transformation $x \rightarrow y$ incurs information loss. Let us formulate the additional latent variables as $z \in \mathbb{R}^D$ such that for the INN shown in Figure 1, the dimensionality of $[y, z]$ is equal to the dimensionality of $x$ or $M + D = N$. Hence, in the inverse process, the objective is to deduce the high-dimensional conductivity $x$, from a sparse set of magnetic field measurements $y$. Note that $x$ can be either the current distribution or the conductivity map, where the former was difficult to reconstruct based on the INNs.

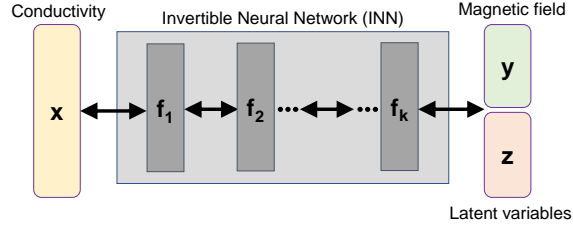

**Figure 1. An overview of our INN architecture**

## 2 METHOD

### 2.1 Forward Problem - Biot-Savart Law

We define the conductivity map $x$ as $\sigma$ and the applied electric field in 3D space as $E$. Since neither the liquid metal nor the non-conducting bubble void fractions inside the conductor in our simulation setup (Krause et al., 2023) are moving, the Ohm's Law at a position $r$ results in $j(r) = \sigma(r)E(r)$ where $j(r)$ is the current density. With the known current density at pre-defined points, the induced magnetic flux density at a point $r$ in 3D space is computed using the Biot-Savart law,

$$B(r) = \frac{\mu_0}{4\pi} \int_V \frac{j(r') \times (r-r')}{|r-r'|^3} dV \qquad (1)$$

$\mu_0$ is the permeability of free space, $V$ is the volume with $dV$ as infinitesimal volume element and $B(r) \in \mathbb{R}^3$ is the magnetic field at point $r$. We term the measurable component of $B(r)$ as $y(r)$ while $r'$ is the integration variable and a location in $V$. In (1) and our simulation, steady-state current flow is assumed. But, for time-varying current or magnetic field, the time derivative of the fields must be considered.

### 2.2 Invertible Neural Network (INN)

The overview of our INN model is provided in Figure 1 which closely follows (Ardizzone et al., 2019). The INN has a bijective mapping between $[y, z]$ and $x$, leading to INN's invertibility, that learns to associate the conductivity $x$ with unique pairs $[y, z]$ of magnetic field measurements $y$ and latent variables $z$. We incorporate the latent variables $z$ to address the information loss in the forward process $x \to y$. Assuming INN is an invertible function $f$, the optimization via training explicitly calculates the inverse process, i.e. $x = f(y, z; \theta)$ where $\theta$ are the INN parameters. We define the density of the latent variable $p(z)$ as the multi-variate standard Gaussian distribution. The desired posterior distribution $p(x|y)$ can now be represented by the deterministic function $f$ that pushes the known Gaussian prior distribution $p(z)$ to $x$-space, conditioned on $y$. Note that the forward mapping, i.e. $x \to [y, z]$ and the inverse mapping, i.e. $[y, z] \to x$ are both differentiable and efficiently computable for posterior probabilities. Therefore, we aim to approximate the conditional probability $p(x|y)$ by our tractable INN model $f(y, z; \theta)$ which uses the training data $\{(x_i, y_i)\}_{i=1}^T$ with $T$ samples from the forward simulation.

### 2.3   INN Architecture and Training Loss

Our INN model $f$ is a series of $k$ invertible mappings called coupling blocks with $f \coloneqq (f_1, \ldots, f_j, \ldots, f_k)$ and $x = f(y, z; \theta)$. Our coupling blocks are learnable affine transformations, scaling $s$ and translation $t$, such that these functions need not be invertible and can be represented by any neural network (Dinh et al., 2017). The coupling block takes the input and splits it into two parts, which are transformed by $st$ networks alternatively. The transformed parts are subsequently concatenated to produce the block's output. The architecture allows for easy recovery of the block's input from its output in the inverse direction, with minor architectural modifications ensuring invertibility. We follow (Kingma and Dhariwal, 2018) to perform a learned invertible $1 \times 1$ convolution after every coupling block to reverse the ordering of the features, thereby ensuring each feature undergoes the transformation. Even though our INN can be trained in both directions with losses $L_x$, $L_y$ and $L_z$ for variables $x$, $y$, $z$ respectively as in (Ardizzone et al., 2019), we are only interested with reconstructing the conductivity variable $x$, i.e. the inverse process. Additionally, leaving out $L_y$ and $L_z$ allows us to not perform the manual optimization of the weights of multiple loss definitions for stable training. Given the batch size as $W$, the loss $L_x$ minimizes the reconstruction error between the groundtruth and predictions during training as follows:

$$L_x(\theta) = \left(\frac{1}{W}\sum_{i=1}^{W}|x_i - f(y_i, z_i, \theta)|^2\right)^{\frac{1}{2}} \quad \text{with objective} \quad \theta^* = \text{argmin}_\theta L_x(\theta) \qquad (2)$$

## 3   EXPERIMENTS AND RESULTS

### 3.1   Simulation setup and Data pre-processing

We calculate the dataset for training and testing our INN model by using the proof-of-concept (POC) simulation setup by (Krause et al., 2023) as shown in Figure 2 (left). The model simplifies the water electrolyzer to an electrical conductor with dispersed non-conducting components. Through Cu wires, with (length, width, height) of 50 x 0.5 x 0.5 cm, connected to Cu electrodes (10 x 7 x 0.5 cm), a current is applied to liquid GaInSn. The liquid metal is filled into a thin channel of (16 x 7 x 0.5 cm), and thereafter, the conductive electrolyte is simulated. With the use of GaInSn, reactions at the electrode surfaces and, thus, concentration-induced conductivity gradients are excluded.

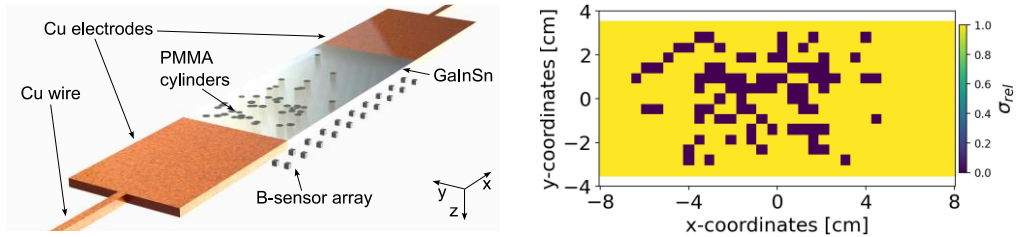

**Figure 2.** The POC model with Cu electrodes and wire, liquid GaInSn with PMMA cylinders and magnetic sensors (left). The binarized conductivity distribution of GaInSn containing region shown in x-y Cartesian plane (right).

The gas bubbles are simulated in the quasi-two-dimensional setup by using between 30 and 120 PMMA cylinders with diameters from 4 to 5 mm and negligibly low electrical conductivity placed in GaInSn. To measure the magnetic flux density field, an array of 10 x 10 sensors is positioned at a distance ($d_{sensor}$) of 5 mm and 25 mm below the electric current carrying part of the setup that contains GaInSn. In our future experimental setup, only one spatial component of the magnetic flux density is measurable. Hence, the conductivity map, together with one spatial component of the magnetic flux density, act as the groundtruth data. We simulated the electric conductivity distributions of 10,000 different geometrical configurations with a fixed applied current strength. After transforming $\sigma$ of the initial variously dimensioned tetrahedral to a hexahedral mesh with defined dimensions, the resulting conductivities were divided with $\sigma_{GaInSn} = 3.3 \cdot 10^6$ S/m, giving $\sigma_{rel}$ between 0 and 1. Subsequently, the relative conductivities were binarized by assigning values smaller than 0.25 to 0 and equal or superior to 1. A binary conductivity map of a sample is shown in Figure 2 (right). More details related to our simulation setup can be found in (Krause et al., 2023). From the originally 774 simulated conductivity data points, we selected only those directly above the sensor positions, resulting in 510 data points. Hence, for INN training, the data comprises magnetic field values with 100 sensor features and a conductivity map with 510 features for each geometry. To create training and validation sets, we shuffled the geometries and allocated 80% for training and 20% for validation. Standardization of the data was performed to ensure a common scale and distribution of conductivity and magnetic field features.

### 3.2 Comparison with classical approaches

We obtained qualitative results of our INN model and compared it with regularization approaches (Tikhonov and ElasticNet) in Figure 3. The evaluation was performed using data with a sensor distance ($d_{sensor}$) of 5 mm and 100 sensors. The regularization parameters for Tikhonov and ElasticNet were determined through cross-validation on the training set. Our INN model shows a good approximation of the groundtruth, providing meaningful insights into the location of the non-conducting PMMA cylinder-induced void fractions mimicking the bubbles. The results of Tikhonov and ElasticNet regularization were similar, indicating the minimal impact of the $L_1$ penalty in ElasticNet for improving the predictions.

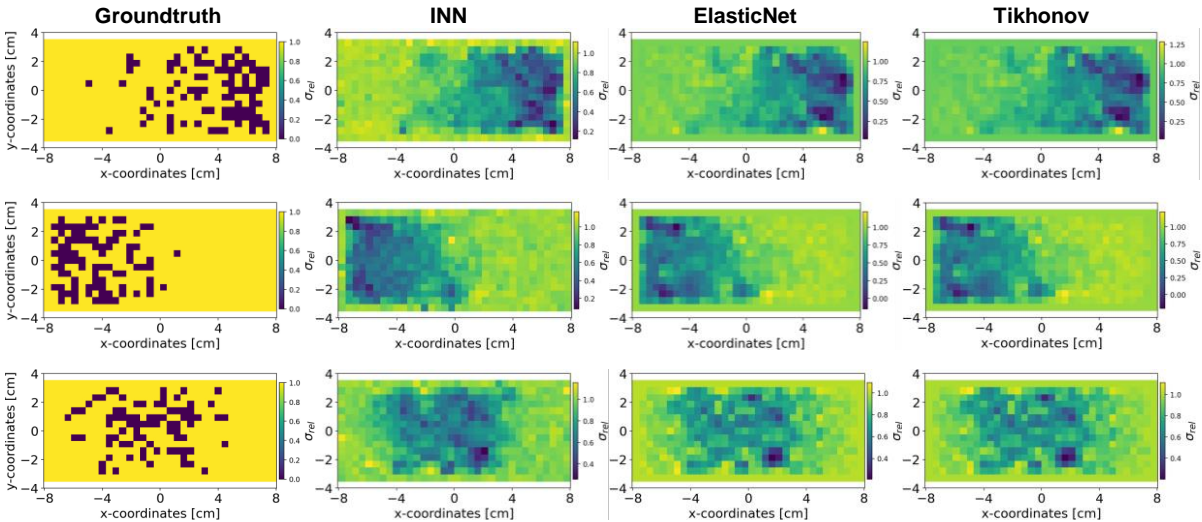

**Figure 3. The results from different reconstruction models for data with $d_{sensor}$ = 5 mm and 10 x 10 sensors.**

We also compared the latency for fitting ElasticNet and Tikhonov models and training our INN with four coupling blocks on similar hardware. The ElasticNet took approximately 4 hours, the Tikhonov took 45 minutes, while our INN training took only 142 seconds on a single GPU. Note that the reported time for ElasticNet and Tikhonov models includes the regularization parameter tuning process. These timings present a significant speed advantage of our INN model compared to the other approaches.

### 3.3 Ablation study on $d_{sensor}$ and number of sensors

We performed an ablation study to investigate the effect of changing the distance of the sensors from the conducting plate and the number of sensors. Figure 4 shows the results obtained after training separate instances of our INN model. Interestingly, the INN can reconstruct the placement of PMMA cylinder-induced void fractions even in the simulation setup with only 50 sensors and a sensor distance of 25 mm. However, the pixel-level correlations with adjacent data points are slightly degraded. Since statistical models, including INN, provides continuous valued predictions, we quantitatively evaluate the performance of our INN-based approach with those from classical approaches in Section 3.5.

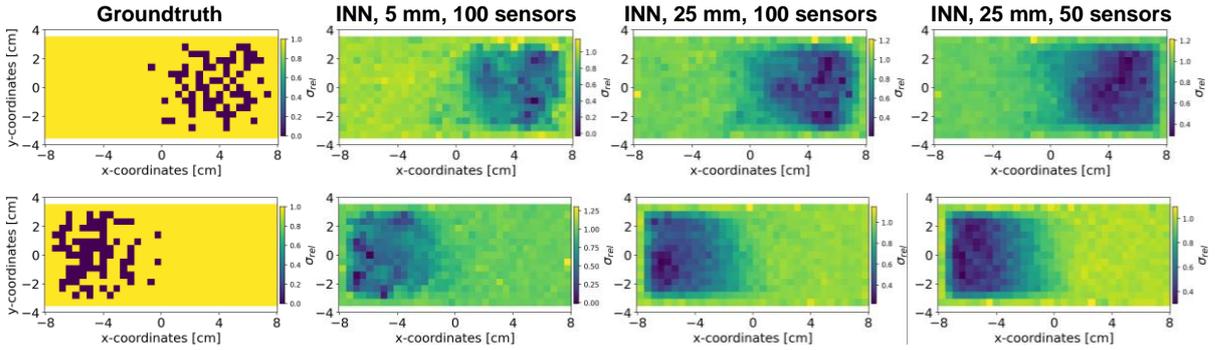

**Figure 4. The results from our INN model after varying distance of sensors from the plate and number of sensors.**

### 3.4 Validation Loss vs Coupling Blocks (k)

Multiple instances of the INN model were trained on various experimental settings with batch size 100, learning rate $\alpha = 1e-4$, and Adam Optimizer ($\beta_1 = 0.8$ and $\beta_2 = 0.9$). Figure 5 (top row) displays validation loss curves for different numbers of coupling blocks (*k*) in the INN. Using only one coupling block leads to underfitting, while a higher number of blocks can cause overfitting. We stop training when the validation loss begins to increase. Notably, increasing *k* beyond three does not significantly reduce the validation loss, making it difficult to determine the best convergence. For the setup with 25 mm distance and 100 sensors, validation losses are higher compared to the setup with 5 mm distance and 100 sensors due to reduced information in magnetic field measurements with greater sensor distance. The setup with 25 mm distance and 50 sensors further degrades information. However, Figure 4 demonstrates the INN's ability to learn the PMMA cylinder distribution. Validation loss scores at the last epoch in Figure 5 reveal higher loss values for greater sensor distance and fewer sensors compared to the setup with 5 mm and 100 sensors, and the optimal *k* for coupling blocks to be 3 to 4 for each setup.

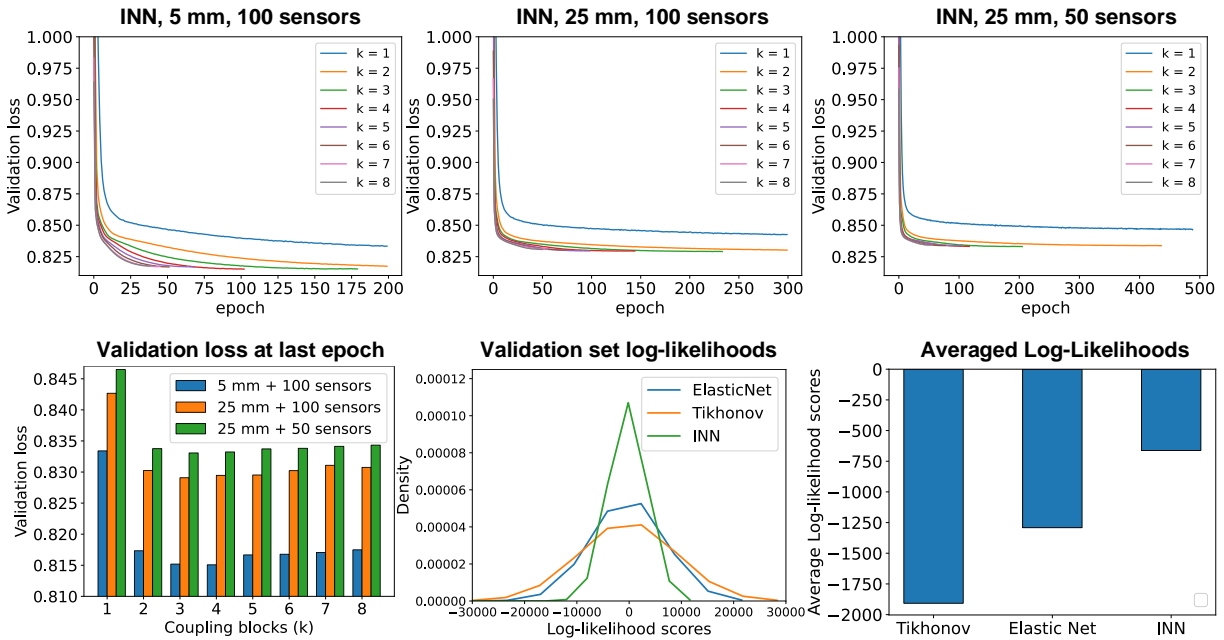

**Figure 5. Top row shows the validation losses with varying INN coupling blocks. The centre image in the bottom row shows the log-likelihoods of the groundtruth with respect to the probability distribution of binary ensemble maps via error diffusion. The right image in the bottom row are the averaged log-likelihoods through the entire validation set.**

### 3.5 Random Error Diffusion

As the visual comparison in Figure 3 is not conclusive to determine the best reconstruction approach, we developed an ensemble-based evaluation to convert continuous maps to discrete conductivity values. In principle, Floyd-Steinberg dithering (known as Error Diffusion) can be used, but it spreads quantization errors into neighboring pixels with pre-defined fractions, due to which this technique won't reproduce exact binary groundtruth. We, therefore, randomize error fractions using Dirichlet distribution and generate an ensemble of binary maps from each predicted continuous conductivity map. Note that the fractions sum up to 1, and for computation constraints, we generate an ensemble of 100 binary maps. Next, the probabilistic density of the binary ensembles is estimated for each groundtruth, and the likelihood of the groundtruth with respect to the estimated density is computed. Figure 5 (bottom center) displays log-likelihood scores as a kernel density plot for validation samples (setup: 5 mm, 100 sensors). The Tikhonov model shows greater deviation compared to ElasticNet, whereas our INN model exhibits the least deviation, as confirmed by the averaged log-likelihood scores in Figure 5 (bottom right). Hence, the INN provides higher likelihood scores for the groundtruth compared to other approaches.

## 4 CONCLUSION

In this study, we proposed Invertible Neural Networks (INNs) to reconstruct conductivity maps from external magnetic field measurements in a model simulation setup mimicking features of a water electrolyzer. The results demonstrate the robustness of our INN model in learning the conductivity distribution, despite the ill-posed nature of the problem. Quantitative evaluation using randomized error diffusion confirms that INN provides accurate conductivity map approximations and significantly improves the likelihood that the predictions resemble the groundtruth. Our findings show that INNs can effectively reconstruct conductivity maps with a low number of sensors and at distances greater than 20 mm. Hence, INNs offer a promising approach for localizing and estimating non-conductive fractions in current conducting liquids, with potential for practical applications. Future research directions include investigating INN performance on higher-resolution conductivity maps and performing experiments with sensor measurements that contain noisy readings.